\documentclass[prd, amsfonts, twocolumn, nofootinbib, showpacs]{revtex4}
\usepackage{graphicx, epsfig}
\usepackage{color}
\usepackage{amsmath}
\usepackage{hyperref}
\usepackage{subfig}
\usepackage{tabularx}
\usepackage{amsmath,amssymb}
\newcommand{\be}{\begin{equation}}
\newcommand{\ee}{\end{equation}}
\newcommand{\bea}{\begin{eqnarray}}
\newcommand{\eea}{\end{eqnarray}}

\newcommand{\gapp}{\mathrel{\raise.3ex\hbox{$>$}\mkern-14mu \lower0.6ex\hbox{$\sim$}}}
\newcommand{\lapp}{\mathrel{\raise.3ex\hbox{$<$}\mkern-14mu \lower0.6ex\hbox{$\sim$}}}
\def\bbox{{\,\lower0.9pt\vbox{\hrule \hbox{\vrule height 0.2 cm
\hskip 0.2 cm \vrule  height 0.2 cm}\hrule}\,}}

\usepackage{color}
\usepackage{ulem}
\usepackage[usenames,dvipsnames]{xcolor}

\begin{document}
\title{Gravity can significantly modify classical and quantum Poincar\'e recurrence theorems}
\author{Ruifeng Dong and Dejan Stojkovic}
\affiliation{ HEPCOS, Department of Physics, SUNY at Buffalo, Buffalo, NY 14260-1500}

\begin{abstract}
\widetext
Poincar\'e recurrence theorem states that any finite system will come arbitrary close to its initial state after some very long but finite time. At the statistical level, this by itself does not represent a paradox, but apparently violates the second law of thermodynamics, which may lead to some confusing conclusions for macroscopic systems. However, this statement does not take gravity into account. If two particles with a given center of mass energy come at the distance shorter than the Schwarzschild diameter apart, according to classical gravity they will form a black hole. In the classical case, a black hole once formed will always grow and effectively quench the Poincar\'e recurrence. We derive the condition under which the classical black hole production rate is higher than the classical Poincar\'e recurrence rate. In the quantum case, if the temperature of the black hole is lower than the temperature of the surrounding gas, such a black hole cannot disappear via Hawking evaporation.
We derive the condition which gives us a critical temperature above which the black hole production is faster than quantum Poincar\'e recurrence time. However, in quantum case, the quantum Poincar\'e recurrence theorem can be applied to the black hole states too. The presence of the black hole can make the recurrence time longer or shorter, depending on whether the presence of the black hole increases or decreases the total entropy. We derive the temperature {\it below} which the produced black hole increases the entropy of the whole system (gas particles plus a black hole). Finally, if evolution of the system is fast enough, then newly formed black holes will merge and accrete particles until one large black hole dominates the system. We give the temperature {\it above} which the presence of black holes increase the entropy of the whole system and prolongs the Poincar\'e recurrence time.
\end{abstract}


\pacs{}
\maketitle

{\it Introduction.~}
The second law of thermodynamics implies that the total entropy of a closed system can never decrease. As a consequence, spontaneous processes are irreversible, which introduces the asymmetry between future and past, i.e. the arrow of time.  If the systems starts from some arbitrary state of low entropy, it will evolve towards the equilibrium which is by definition a state of the highest entropy. The second law of thermodynamics will then forbid any further interesting evolution. However, any system can and will undergo fluctuations. After long enough time, very large (and thus unlikely) fluctuations will happen which can take the system far from the equilibrium, and possibly back to the original  state with very low entropy. This Poincar\'e recurrence time is exponential in entropy of the system in the classical case, and doubly exponential in the quantum case, which is unfathomably large for any reasonable macroscopic system. However, the mere fact that the second law of thermodynamics is violated in this process of recurrence can lead to some puzzling consequences like the "Boltzmann brain problem" \cite{Page:2006hr,Albrecht:2004ke,Boddy:2015fqa,Carlip:2007id}.

In this paper we point out that usual discussion of the Poincar\'e recurrence problem does not take gravity into account. However gravity can significantly modify behavior of a thermodynamical system. Black holes can be classically produced in energetic collisions of particles with trans-Planckian center of mass energy.  In the classical case, according to the Hawking area theorem, a black hole once formed can only grow and its entropy will always increase. Thus, a system containing an appropriate classical black hole (more massive than the Planck mass) could effectively stop the Poincar\'e recurrence from happening.  The events of classical black hole production in collisions of particles are also rare (since the center of mass energy must be trans-Planckian while the temperature for the gas is usually much lower), so the crucial question is which process dominates. We calculate the rate at which such black holes are produced in collisions of particles in a closed system, and find the condition for which this rate is higher than the Poincar\'e recurrence rate.

In the quantum case, two things change. First, black holes can evaporate, and second, the Poincar\'e recurrence time is a double exponential of an entropy of the system. If the Hawking temperature of a black hole is lower than the temperature of its environment, then accretion wins over evaporation. Such a black hole can only grow and its entropy will always increase. However, in quantum case the Poincar\'e theorem must be applied to the black hole itself.  The presence of the black hole can make the Poincar\'e recurrence time longer or shorter, depending on whether the presence of the black hole increases of decreases the total entropy. We derive the condition for which a surviving black hole can be produced, as well as the dividing temperature between the increase and decrease in the Poincar\'e recurrence time due to the presence of the black holes.

{\it Black hole production in gas.~}
We consider an isolated system of ideal gas consisting of particles of the same species and mass m, within a finite volume V. In equilibrium, the particles are uniform in space, and follow a specific momentum distribution. From isotropy, this distribution function only depends on energy of the particle. We write it as $f(E)$. In the following, physical constants will be written explicitly, including the Planck constant $h$, the reduced Planck constant $\hbar$, the Newton's gravitational constant $G$, the Boltzmann constant $k_B$ and the speed of light $c$.

The effective cross section for black hole production in collision of two particles \cite{Dai:2007ki}, labeled as 1 and 2, is
\be \label{gcs}
\sigma_{12}=\pi(2R_S)^2,
\ee
where $R_S=2GM/c^2$ is Schwarzschild radius for the center-of-mass energy $Mc^2$.
The two particles have momenta $\vec{k}_1$, $\vec{k}_2$ , and energies $E_1$, $E_2$ before collision, so
\bea\label{Mass}
M&=&\frac1{c^2}\sqrt{(E_1+E_2)^2-c^2(\vec{k}_1+\vec{k}_2)^2}\nonumber\\
&=&\sqrt{2m^2+2E_1E_2/c^4-2k_1k_2\cos\theta/c^2},
\eea
where $\theta$ is the angle between $\vec{k}_1$ and $\vec{k}_2$. Here, we used the energy-momentum relations
\be
E_1^2=k_1^2c^2+m^2c^4, ~~~~E_2^2=k_2^2c^2+m^2c^4.
\ee

With these preliminaries, we are ready to calculate the collision rate. Without loss of generality, we single out a particle 1. During some short time interval $dt$, the expected number of collisions experienced by particle 1 with particles with all possible momenta
$\vec{k}_2$ is\footnote{Here we assumed the system is not too dense, so that the particle's mean free path is much larger than the average Schwarzschild radius
for a typical collision. More precisely, we mean that $\langle{n_{gas}}\rangle\langle{2R_S}\rangle^3\ll 1$, where the $\langle\rangle$ means the average over all particles and collisions. This means, approximately, the temperature is lower than the Planck temperature,
as will be quantified later.},
\be
dN^{(1)}_{col}=\bar{n}dt\int d^3k_2 v_{12}(\vec{k}_1,\vec{k}_2)\sigma_{12}(\vec{k}_1,\vec{k}_2)f(E_2),
\ee
where $\bar{n}$ is the uniform number density and $v_{12}$ is the magnitude of the relative velocity between the two particles, which can be calculated as
\bea
v_{12}(\vec{k}_1,\vec{k}_2)&=&|\vec{v}_1-\vec{v}_2| \nonumber\\
&=&\sqrt{v_1^2+v_2^2-2v_1v_2\cos\theta} \nonumber\\
&=&\sqrt{(\frac{c^2k_1}{E_1})^2+(\frac{c^2k_2}{E_2})^2-2\frac{c^4k_1k_2}{E_1E_2}\cos\theta}.\nonumber\\
\eea
Here, we used the relation between velocity and momentum, i.e. $\vec{v}_{i}=\frac{c^2\vec{k}_i}{E_i}$ for $i=1,2$. 

Note the relative velocity $v_{12}$ here is used to convert the cross section (reaction rate per flux) to reaction rate per density. Its form applies to 
relativistic collisions, as it makes the following production rate in Eq. (\ref{BHrate0}) covariant.\cite{Weinberg:1995mt}

The total collision rate per unit volume, which is also the expected black hole formation rate, is
\small
\bea\label{BHrate0}
\frac{dn_{BH}}{dt}&=& \frac12 \bar{n} \int d^3k_1 f(E_1) \frac{dN^{(1)}_{col}}{dt} \nonumber\\
&=& \frac12\bar{n}^2\iint d^3k_1d^3k_2 v_{12}(\vec{k}_1,\vec{k}_2)\sigma_{12}(\vec{k}_1,\vec{k}_2)f(E_1)f(E_2),\nonumber\\
\eea
\normalsize
where the $\frac12$ factor comes from the interchangeability of particles 1 and 2.

To simplify the calculations, from now on, we concentrate on ultrarelativistic particles with $m\ll k_BT/c^2$.
Later, we will evaluate the above integral with the constraint $M\ge M_P$ for the
 classical case and $T_{BH}\le T$ for the quantum case. Here $M_P=\sqrt{\hbar c/G}$ is the Planck mass and $T_{BH}=\frac{\hbar c^3}{8\pi GMk_B}$ is the Hawking temperature of
 the black hole.

{\it Classical system.~}
We first start with the classical system. The momenta of particles follow the Maxwell-J\"uttner distribution \footnote{Strictly, the microcanonical ensemble is used to describe isolated system with a fixed total energy. But the two-particle collision we are considering happens in a region much smaller than the size of the macroscopic system. So for this system in thermal equilibrium, we can use the canonical ensemble description, that is, considering the volume consisting of much smaller patches, each of which is in thermal contact with the rest of the system.}, i.e.
\be
f_c(E)=\frac1{4\pi m^3c^3\Theta K_2(1/\Theta)}\exp(-E/k_BT),
\ee
where $\Theta=\frac{k_BT}{mc^2}$ and $K_n(x)$ is the nth-order modified Bessel function of the second kind. The subscript $c$ stands for classical.

The particles are ultrarelativistic in our consideration, so we set $m=0$ for simplicity. Define two dimensionless vectors,
\be\label{x1x2}
\vec{x}_1\equiv\frac{c\vec{k}_1}{k_BT}, ~~~~\vec{x}_2\equiv\frac{c\vec{k}_2}{k_BT},
\ee
Eq. (\ref{BHrate0}) can be decomposed into a dimensional P and a dimensionless term A as
\be\label{BHrate1}
\frac{dn_{BH}}{dt}= P\times A.
\ee
The two terms are, respectively,
\bea\label{PA}
P&\equiv&\frac{\sqrt{2}G^2(k_BT)^2\bar{n}^2_c}{4\pi c^{13}}\nonumber\\
A&\equiv&\iint d^3x_1d^3x_2 \frac{x_1x_2(1-\cos\theta)^{3/2}}{\exp(x_1+x_2)}.
\eea
Here $\theta$ is the angle between dimensionless vectors $\vec{x}_1$ and $\vec{x}_2$. We have taken $K_2(1/\Theta)=2\Theta^2$ in the limit $\Theta\rightarrow \infty$.

A black hole can be treated as classical if its mass is larger than the Planck mass, i.e. $M\ge M_P$. Once this black hole is formed, it will stay in the system and accrete
surrounding particles. In this classical case black holes do not evaporate (in the quantum case we will let the black holes evaporate).

If the formation rate of surviving black holes is higher than the Poincar\'e recurrence rate, the entropy of the whole system
would partly remain in the form of black hole entropy, which never decreases according to the Hawking area theorem. If this is the case, the total entropy would not
be able to decrease to become arbitrarily small as predicted by the Poincar\'e recurrence theorem.

Using Eqs.~(\ref{Mass}) and (\ref{x1x2}), $M\ge M_P$ can be written as a constraint for the integral (\ref{PA}), i.e.
\be\label{constraint1}
x_2\ge\frac{F(\theta)}{x_1}.
\ee
Here $F(\theta)=\frac1{2\tilde{T}^2(1-\cos\theta)}$. We introduced the dimensionless temperature
$\tilde{T}\equiv\frac{T}{T_P}$, where $T_P\equiv \sqrt{\frac{\hbar c^5}{Gk_B^2}}$ is the Planck temperature.

The radial part of A under this constraint is then
\bea
&&\int_0^{\infty}dx_1\frac{x_1^3}{\exp(x_1)}\int_{F(\theta)/x_1}^{\infty}dx_2 \frac{x_2^3}{\exp(x_2)}\nonumber\\
&=&\int_0^{\infty}dx_1\frac{F^3+3F^2x_1+6Fx_1^2+6x_1^3}{\exp(x_1+F/x_1)}\nonumber\\
&=&2F^{5/2}(3+F)K_3(2F^{1/2}) + 2F^{2}(6+F)K_4(2F^{1/2}).\nonumber\\
\eea
In the second step, we used the integral representation and recurrence relations for the modified Bessel functions
\footnote{We used Eqs.~(10.29.1) and (10.32.10) of this reference.} \cite{Olver:2010:NHMF}.

Functions $K_{3,4}(x)$ are both monotonically decreasing functions with the positive argument $x$. So their lower limit can be
provided by the asymptotic approximations as $x\rightarrow +\infty$.
\be
K_{3,4}(x)>\sqrt{\frac{\pi}{2x}}\exp(-x).
\ee
We define $\lambda\equiv\frac{\sqrt{2}}{\tilde{T}}$, so $F(\theta)=\frac{\lambda^2}{4(1-\cos\theta)}$. The integration in $A$ is then
\small
\bea\label{Alower}
A&\gtrsim&\sqrt{\frac{\pi}{512}}\lambda^{7/2}\iint d\Omega_1d\Omega_2\Big[\frac{\lambda^3}{4}(1-\cos\theta)^{-7/2}\Big.\nonumber\\
&&\Big.+\frac{\lambda^2}{2}(1-\cos\theta)^{-3}+ 3\lambda(1-\cos\theta)^{-5/2}+12(1-\cos\theta)^{-2} \Big]\nonumber\\
&&\times(1-\cos\theta)^{1/4}\exp\big(-\lambda(1-\cos\theta)^{-1/2}\big).
\eea
\normalsize

We also define
\be
g(\lambda)\equiv\iint d\Omega_1d\Omega_2(1-\cos\theta)^{1/4}\exp\big(-\lambda(1-\cos\theta)^{-1/2}\big).
\ee
Evaluation of this function will be given in App.~(\ref{g(lambda)}), and we just show the result here.
\small
\bea
g(\lambda)&=&\frac{2^{25/4}\pi^2}{15} \bigg[\left(2 \lambda ^2-\sqrt{2} \lambda +3\right)\exp\left(-\frac{\lambda }{\sqrt{2}}\right)\bigg.\nonumber\\
&&~~~~~~~~~~~~~~~~~~~\bigg.-2^{3/4} \sqrt{\pi } \lambda ^{5/2} \text{erfc}\bigg(\frac{\sqrt{\lambda }}{\sqrt[4]{2}}\bigg)\bigg],
\eea
\normalsize
where $\text{erfc}(x)$ is the complementary error function.

The lower limit of $A$ can be written as a weighted sum of derivatives of $g(\lambda)$ with $\lambda$, and then as a function of the dimensionless temperature $\tilde{T}$.
\bea\label{Afinal}
A&\gtrsim&\sqrt{\frac{\pi}{512}} \lambda ^{7/2} \bigg[-\frac{1}{4} \lambda ^3 \text{g}^{(7)}(\lambda )+\frac{1}{2} \lambda ^2 \text{g}^{(6)}(\lambda )\bigg.\nonumber\\
&&~~~~~~~~~~~~~~~\bigg.-3 \lambda  \text{g}^{(5)}(\lambda )+12 \text{g}^{(4)}(\lambda )\bigg] \nonumber\\
&=&\frac{\pi ^{5/2}}{2^{11/2} \tilde{T}^{11/2}} \left[693 \sqrt{\pi } \tilde{T}^{7/2} \text{erfc}\left(1/\tilde{T}^{1/2}\right)\right.\nonumber\\
&&\left.+\left(1386 \tilde{T}^3+412 \tilde{T}^2+88 \tilde{T}+16\right)e^{-1/\tilde{T}}\right].\nonumber\\
\eea
Using Eqs.~(\ref{BHrate1}) and (\ref{Afinal}), we are ready to write down the expected formation rate of a black hole which will survive and grow.
\small
\bea\label{BHratefinal}
r_{BH,c}&=&V\frac{dn_{BH}}{dt}\nonumber\\
&\gtrsim&\frac{\pi^{3/2}\tilde{V}\tilde{\bar{n}}^2_c}{128\tilde{T}^{7/2}} \left[693 \sqrt{\pi } \tilde{T}^{7/2} \text{erfc}\left(1/\tilde{T}^{1/2}\right)\right.\nonumber\\
&&\left.+\left(1386 \tilde{T}^3+412 \tilde{T}^2+88 \tilde{T}+16\right)e^{-1/\tilde{T}}\right]t_P^{-1}.\nonumber\\
\eea
\normalsize
where $t_P=\sqrt{\hbar G/c^5}$ is the Planck time, $\tilde{V}\equiv V/V_P$ is the dimensionless volume and $\tilde{\bar{n}}_c=\bar{n}_cV_P$, with $V_P=\sqrt{\hbar^3G^3/c^9}$ being the Planck volume.

{\it Classical Poincar\'e recurrence.~}
Classical Poincar\'e recurrence theorem states that any isolated finite mechanical system will return to a state infinitely close to its initial state after
 a long but finite time. The finite size of the system is necessary to make the recurrence time finite since in this case the phase space and thus the number of states is finite. This time is called the Poincar\'e recurrence time, which according to \cite{Susskind:2003kw,Dyson:2002pf} is
\be\label{tr}
t_{re,c}=\exp(S/k_B)t_P,
\ee
where $S$ is the total entropy of the system, which can be calculated using the Gibbs' formula \cite{Landau:1980mil},\footnote{The entropy of this system is defined as the sum of the entropy of all its subsystems. Each subsystem (smaller but macroscopic)
can be described by a canonical ensemble. Therefore the Gibbs' formula can be used here.\cite{Landau:1980mil}}
\small
\bea\label{S_classical}
S/k_B&=&-\int\frac{f^N(E)}{V^N}\ln\frac{h^{3N}f^N(E)}{V^N}d^{3N}kd^{3N}y-\ln N!\nonumber\\
&\approx&-N\left(\int f^N(E)\ln\frac{h^3f_c(E)}{V}d^{3N}k+\ln N-1\right)\nonumber\\
&\approx&-N\left(\ln\frac{h^3c^3}{8\pi V(k_BT)^3} - \bigg\langle\left(\frac{E}{k_BT}\right)^{1/N}\bigg\rangle^N+\ln N-1\right)\nonumber\\
&\approx&N\left(\ln\frac{(k_BT)^3}{\pi^2\hbar^3c^3\bar{n}_c}+\exp(3/2-\gamma)+1\right)\nonumber\\
&\approx&N\ln\frac{3.41(k_BT)^3}{\hbar^3c^3\bar{n}_c}.
\eea
\normalsize
The coordinates of particles are labeled by $y$, and $\gamma$ is the Euler constant. In the first line, we used the fact that $\frac{f^N(E)}{V^N}$ is the phase-space
distribution function. The $h^{3N}$ factor converts the classical phase-space volume to the number of states. The $\ln N!$ provides the correct Boltzmann counting.
This factor is evaluated using the Stirling's approximation in the second line. In the third line, we have taken the limit $mc^2/k_BT\rightarrow 0$ to evaluate
$K_2(1/\Theta)$. Also in this limit, the average of $\left(\frac{E}{k_BT}\right)^{1/N}$ is evaluated in the following line. In the last line, all numerical
quantities are put together inside the logarithmic function. As expected, entropy is now an extensive state function.

The classical Poincar\'e recurrence time is therefore
\be
t_{re,c}=\left(3.41\frac{\tilde{T}^3}{\tilde{\bar{n}}_c}\right)^N t_P.
\ee

In general, one might question validity of the classical Poincar\'e recurrence scale
for a generic chaotic many body system. Given any a priori initial resolution for the initial state in classical
phase space, the time evolution will conserve volume but will make
the region look convoluted and fractal in phase space, thus complicating its local
structure. The phase space distribution will acquire structure on the level of the
quantum phase space resolution, or order $\hbar$. Beyond this point, the detailed continuous
classical evolution makes no physical sense. Note however, that in calculation of the classical Gibbs entropy in Eq.~(\ref{S_classical}) one indeed uses
the $h^{3N}$ factor to convert the classical phase-space volume to the number of states. It is this entropy that enters the expression for the recurrence time in Eq.~(\ref{tr}).
Thus, the expression Eq.~(\ref{tr}), which is usually used in the literature \cite{Susskind:2003kw,Dyson:2002pf}, in this context is appropriate.

{\it Comparing BH formation rate and recurrence rate.~}
Define now a parameter
\be
\eta_c\equiv r_{BH,c}t_{re,c}.
\ee
If $\eta_c\gg 1$, black holes are produced well before $t_{re_c}$, and the Poincar\'e recurrence can be avoided. Then we expect more and more black holes
to be formed in collisions until the condition $\eta_c\gg 1$ is no longer satisfied. This will happen when black holes (or perhaps a single large black hole made in mergers of the small ones) dominate the system and there are few remaining gas particles in the system.

Now, $\eta_c$ can be written as
\be
\eta_c=\frac{N^2}{\tilde{V}\tilde{T}^{7/2}}\exp\left(N\ln\frac{3.41\tilde{V}\tilde{T}^3}{N}-\frac{1}{\tilde{T}}\right)\times(\text{polynomial in }\tilde{T}).
\ee

As the temperature changes, the exponential term changes much faster than the other terms. So the equation $\eta_c=1$ can be approximated by
\be\label{Tsol}
N\ln\frac{3.41\tilde{V}\tilde{T}^3}{N}-\frac{1}{\tilde{T}}=0.
\ee

\begin{figure}[h]
\captionsetup{justification=raggedright,
singlelinecheck=false
}
\includegraphics[width=3.4in]{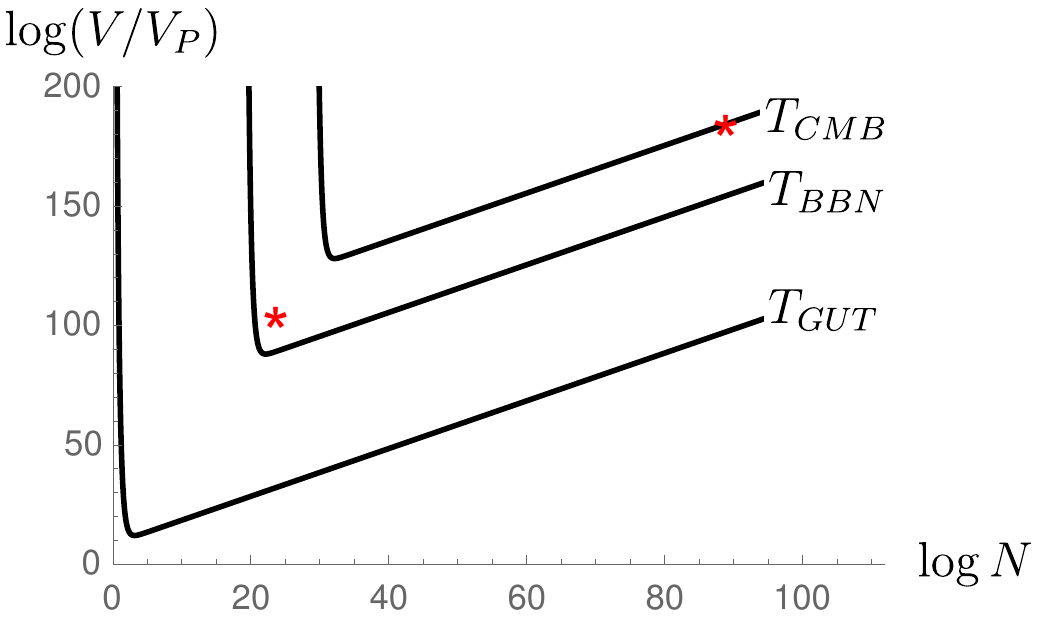}
\caption{$\log(V/V_P)$ as a function of $\log N$, for solutions of Eq. (\ref{Tsol}) with $T=T_{GUT}(\approx10^{16}GeV)$, $T_{BBN}(\approx 1MeV)$, $T_{CMB}(\approx 3K)$. The two red
stars correspond to a typical everyday system with $10^{23}$ particles in $1m^3$ volume (lower left), and the observable universe with $10^{89}$ particles in $4\times 10^{80} m^3$ volume (upper right). Here $10^{89}$ is the number of known particles (mostly photons) in the observable universe, and $4\times 10^{80} m^3$ is this universe's volume. }\label{fig:V-N}
\end{figure}

The solution to this equation gives us a critical temperature  $T=T_C$, above which the black hole production is faster than Poincar\'e recurrence. We can see that, for a fixed number of particles $N$,  larger volume  $V$ implies less frequent particle-particle collision which form black holes, but also gives a longer recurrence due to the larger phase-space volume.
As we can see from Eq.~(\ref{Tsol}), the second effect is more important, so increasing the volume favors the black hole production and works against the Poincar\'e recurrence. Therefore  larger systems correspond to lower $T_C$. On the other hand, for a fixed volume $V$,
a larger $N$ causes more particle collisions and larger phase-space volume, but also a larger correction term in entropy due to Boltzmann counting. As we can see,
the effect of particle collisions is not so important, but the phase-space volume and the correction term in entropy due to Boltzmann counting compete. When $N\ll \tilde{V}\tilde{T}^3$, the phase-space volume is more important, and increasing the number of  particles results in lower $T_C$. When $N\gg \tilde{V}\tilde{T}^3$, the Boltzmann counting term is more important, and increasing the number of  particles causes higher $T_C$. This is clearly seen in Fig. (\ref{fig:V-N}).

In Fig. (\ref{fig:V-N}), we show two particular systems of interest. One is a typical everyday system with the volume of $1 m^3$ and the Avogadro number of particles $(N=10^{23})$. For this system, the critical temperature is $T_C\approx 10^7 K$, which is much higher than room temperature around $300 K$. The other system is our observable
universe, with the total volume of $4\times 10^{80} m^3$ and total number of particles of $10^{89}$. It turns out $T_C\approx T_{CMB} = 3K$, which is the temperature of today's cosmic microwave background (CMB) radiation. If our universe were a system with a fixed volume and number of particles, that would imply that it would never undergo the classical Poincar\'e recurrence. However, since the volume is changing due to expansion, and non-adiabatic processes of particle decay and creation also change the number of particles, one would need to perform a much more careful analysis in order to conclude whether our universe would go through the classical Poincar\'e recurrence.

{\it Quantum systems.~}
In this section we consider the quantum case. We specify an isolated system consisting of massless bosons, which follow the quantum Bose-Einstein distribution in phase space. In momentum space, it has the form
\be
f_q(E)=\frac{g}{h^3 \bar{n}_q[\exp(E/k_BT)-1]},
\ee
where $g$ is number of helicity states, and the subscript $q$ stands for "quantum". Because $\left(\exp(E/k_BT)-1\right)^{-1}>\exp(-E/k_BT)$, we can, as a conservative approximation, still use the same techniques as for the classical system with a few modifications.
The decomposition as in Eq. (\ref{PA}) holds with P changed to
\be
P\rightarrow\frac{16\sqrt{2}\pi g^2 G^2 (k_BT)^8}{h^6 c^{13}}.
\ee

In the semi-classical theory, in addition to accretion black holes can also evaporate. When the Hawking temperature is lower than temperature of the gas, evaporation is faster than accretion, and the black hole survives. So we consider the production rate of a black hole satisfying this condition.
Therefore, the definition of the dimensionless factor A remains the same and can be calculated in the same way, but with a different constraint, i.e. $T_{BH}\le T$. So we have to redefine $F(\theta)$ in Eq. (\ref{constraint1}) as $F(\theta)\rightarrow\frac1{128\pi^2\tilde{T}^4(1-\cos\theta)}$, and then $\lambda$ in Eqs. (\ref{Alower}) to (\ref{Afinal}) as $\lambda\rightarrow\frac1{4\sqrt{2}\pi\tilde{T}^2}$.
We can use this new $\lambda$ in Eq. (\ref{Afinal}), and from the new assignments to factors P and A, we get the black hole formation rate in the quantum case as
\small
\bea\label{r_BH,q}
r_{BH,q}&\gtrsim&\frac{g^2\tilde{V}}{2^{39/2}\pi^{8}\tilde{T}^3}\left[44352\sqrt{2} \pi^4 \tilde{T}^7 \text{erfc}\left(\frac1{2 \sqrt{2 \pi }\tilde{T}}\right)\right.\nonumber\\
&&\left.+(44352 \pi ^3 \tilde{T}^6+1648 \pi ^2 \tilde{T}^4+44 \pi  \tilde{T}^2+1)e^{-\frac{1}{8 \pi  \tilde{T}^2}}\right]t_P^{-1}.\nonumber\\
\eea
\normalsize

Note that we did not have to modify Eq.~(\ref{gcs}) for the black hole production probability.
A black hole of mass $M$ can be viewed as a macroscopic object containing a large
number of micro-states ${\cal N} = \exp{(S_{BH})}$, where the black hole entropy is $S_{BH}= \frac{4\pi G M^2}{\hbar c k_B}$.
The total creation probability of a black hole in collisions of particles is equal to the sum of the probabilities to create a black hole at a given micro-state. Thus,
the total probability in Eq.~(\ref{gcs}) should be multiplied by a pre-factor ${\cal N}$. However, for a given microstate, by the CPT symmetry, we have to add a multiplicative pre-factor which takes into account the probability of the reverse process of the black hole decay into particles. For a highly degenerate system, such as a black hole, the probability to radiate a particle is proportional to the exponent of the corresponding change of the entropy, i.e. $ \exp{\left( -\Sigma_i \Delta S_i\right)}$. Since $\Sigma_i \Delta S_i = S_{BH}$, these two pre-factors cancel each other, and we are left with the original geometric cross-section in Eq.~(\ref{gcs})\cite{Solodukhin:2002ui}.

Since semi-classical black holes have finite number of microstates and entropy, i.e. $S_{BH}\sim \tilde{M}^2\sim\tilde{T}_{BH}^{-2}$, then the quantum Poincar\'e
recurrence can happen to both the gas and the produced black holes.
If black holes are formed at a slower rate than the gas Poincar\'e recurrence, then we expect the recurrence to happen to the gas system, without being affected
by black hole formation. On the other hand, if black holes are formed faster than the gas Poincar\'e recurrence, then one can expect black holes (or perhaps a single large black hole formed in merger of smaller ones) to dominate the system. In that case we have to apply the recurrence theorem to the black hole microstates.

The quantum-mechanical version of the Poincar\'e recurrence time is \cite{Barbon:2003aq}
\be\label{t_re,q}
t_{re,q}=\frac1{\tilde{T}}\exp\left[\exp(S_q/k_B)\ln(1/\epsilon)\right]t_P,
\ee
where $\epsilon$ is the resolution of states in Hilbert space, which is not important for our purpose as shown below.

The total entropy of the gas system using the quantum distribution function is calculated as \cite{weinberg}
\bea\label{entropy}
S_q(T)&=&\frac{V}{T}\int\frac{d^3k_1}{h^3}f_q(E_1)\left(E_1+\frac{c^2k_1^2}{3E_1}\right) \nonumber\\
&=&\frac{4\pi gV}{Th^3}\int_0^{\infty}dk_1\frac{4ck_1^3/3}{\exp(ck_1/k_BT)-1} \nonumber\\
&=&\frac{16\pi^5 gV(k_BT)^3}{45c^3h^3}k_B\nonumber\\
&=&\frac{2\pi^2}{45}g\tilde{V}\tilde{T}^3 k_B.
\eea

Therefore we have
\small
\bea
\eta_q&\equiv& r_{BH,q}t_{re,q}\nonumber\\
&=&\frac{\tilde{V}}{\tilde{T}^4}\exp\left[\exp(\frac{2\pi^2}{45}g\tilde{V}\tilde{T}^3)-\frac1{8\pi\tilde{T}^2}\right]\times\text{(polynomial in }\tilde{T}).\nonumber\\
\eea
\normalsize

As in the classical case, putting the argument of the first exponential to zero yields an approximate solution to $\eta_q=1$, for $\tilde{V}\gg 1$.

\be \label{tprq}
\exp(\frac{2\pi^2}{45}g\tilde{V}\tilde{T}^3)-\frac1{8\pi\tilde{T}^2}=0 .
\ee

As an illustration, consider again a macroscopic system of volume $V\sim 1m^3$. In this case, we get the solution $T_C \sim 10^{-34}T_P\sim 10^{-2} K$. Therefore, this typical macroscopic system at room temperature will experience a period of black hole production before the recurrence happens in the original gas, and then whole system, containing black holes and the remaining gas particles, will undergo Poincar\'e recurrence afterwards. Entropy from the remaining gas and the black holes will then determine the recurrence time in Eq. (\ref{t_re,q}).

Note that here we are considering the production of black hole from particle collisions, not by thermal fluctuations. Thus, we can produce black holes even if this process is not thermally favored, i.e. even if the total entropy in the system decreases due to black hole formation. As a result, the Poincar\'e recurrence time may become shorter or longer.
If a black hole dominates the system and entropy of the remaining gas is negligible, instead of Eq.~(\ref{t_re,q}) the recurrence time is given by
\be
t_{re,BH}=\frac1{\tilde{T}_{BH}}\exp\left[\exp(S_{BH}/k_B)\ln(1/\epsilon)\right]t_P.
\ee

Consider the collision of two particles which form one black hole with mass $M$, Hawking temperature $T_{BH}$ and entropy $S_{BH}$. Right before collision\footnote{For an ideal system of massless bosons, $S\sim VT^3$ (Eq.~(\ref{entropy})), and the energy $E\sim T\langle\bar{n}_q\rangle\sim VT^4$ (Eq. (\ref{n_q})).}, this amount
of energy contributes to the entropy of gas system by $\Delta S_q\sim \tilde{V}\left(\frac{\tilde{M}}{\tilde{V}}\right)^{3/4}$, where $\tilde{M}$ and $\tilde{V}$ are given in Planck units.
After the collision, this amount of the energy of the gas, $\tilde{M}$, goes into a newly formed black hole. Since $\tilde{T}_{BH}^{-1} \sim \tilde{M}$, we get $\Delta S_q\sim  \tilde{V}^{1/4}\tilde{T}_{BH}^{-3/4}$. The condition $S_{BH}\sim\Delta S_q$ then gives $\tilde{T}_{BH}\sim \tilde{V}^{-1/5}$. For a black hole to survive, its temperature must be lower or at least equal to the temperature of the gas, i.e. $T_{BH} \lesssim T$. Since most of the black holes have Hawking temperature around the system temperature, a system with
\be \label{tc}
\tilde{T}\lesssim \tilde{V}^{-1/5}
\ee
would produce black holes that increase the entropy of the whole system (gas particles plus black holes). For example, for $1 m^3$ of volume, we have $T\lesssim 10^{11}K$. This is much higher than the previous $10^{-2}K$, above which black hole production is faster than Poincar\'e
recurrence. Therefore, this system at room temperature would experience Poincar\'e recurrence longer than that expected from the system without gravity, due to black hole production.

Another system of interest is again the whole observable universe, with the volume of $4\times 10^{80} m^3$. From Eq.~(\ref{tprq}) we find the temperature for fast enough black hole production is $T_C \sim 10^{-29}K$, while the upper bound on temperature for increasing the system's recurrence time by black hole production is $10^{-5}K$. These two values are both lower than the CMB temperature. Again, if the universe were a system with constant volume and temperature, the black holes would form and survive, but the Poicar\'e recurrence would become shorter as a result. For more realistic analysis one has to take the time dependence of the relevant quantities in this case.

However, if our universe is truly de Sitter, and the cosmological
constant is indeed constant \cite{Spradlin:2001pw}, then we can make some definite statements.  De Sitter space emits thermal radiation due to the existence of the event horizon, with the temperature given by
\be
T_{dS}=\frac{\hbar c\sqrt{\Lambda}}{2\pi k_B},
\ee
where $\Lambda\approx 10^{-122}l_P^{-2}$ is the cosmological constant. This gives $T_{dS}\approx 10^{-29}K$. After the universe expands for $70$ e-folds from now, the CMB temperature of today's photons will drop below this constant temperature. Afterwards, the observable universe becomes an isolated system at a constant temperature of $10^{-29}$K. On the other hand, the observable volume will still be approximately the same as today's value, since the Hubble parameter is practically constant during the vacuum energy domination.
In this case, from Eq.~(\ref{tprq}) we can calculate the critical temperature needed for production of black holes colder than the environment which will survive. It turns out that this temperature is equal to the  temperature of the de Sitter horizon of $10^{-29}$K  (another coincidence!). Therefore, once the steady state of a truly de Sitter universe is reached, black holes would form and survive, simultaneously increasing the recurrence time due to black hole entropy.

Finally, the above discussion contains a caveat which needs to be addressed. In general, black holes once formed will evolve further by mergers and accretion of
surrounding particles. If accretion is slow enough to be ignored (i.e. slower than the Poincar\'e recurrence time of the gas), then the above discussion stays true. On the other hand, if accretion is fast enough, the upper bound on temperature in Eq.~(\ref{tc}) is no longer relevant. The surviving black hole will accrete mass until it dominates the system. To see how the Poincar\'e recurrence time changes, we compare the total entropy of the gas $S_{q}\sim (\tilde{V}\tilde{T}^3)$ from Eq.~(\ref{entropy}),
with the entropy of the final large black hole after accretion, i.e. $S_{BH}^{final}\sim (\tilde{V}\tilde{T}^4)^2$, since the mass of such black hole contains almost all the initial energy of the gas.  This condition gives
\be \label{tc1}
\tilde{T}\gtrsim\tilde{V}^{-1/5}
\ee
for the initial temperature of the gas which would produce the final large black hole with more entropy than the initial gas and thus increase the recurrence time. Note that this condition is opposite from the  Eq.~(\ref{tc}).

{\it Conclusion.~}
It was recently pointed out that inclusion of gravity can have interesting consequences for simple thermodynamical systems. For example, gravity can impose a maximal temperature that can be achieved in a thermodynamical system without imposing any a priori cut-offs \cite{Dai:2016axz}.
Along the same line, we demonstrated here that gravity can sometimes significantly modify both the classical and quantum versions of the Poincar\'e recurrence theorem.
Creation of black holes in collisions of particles can happen at any temperature. Once formed, a classical black hole can only grow, eventually consuming the whole system. In order to be considered classical at the time of formation, the black hoe mass must be above the Planck mass, which makes this process very suppressed at temperatures below the Planck temperature. However, the Poincar\'e recurrence rate is also exponentially suppressed by the entropy of the system, so the question is which process dominates. In Eq.~(\ref{Tsol})
we derive the condition which gives us a critical temperature above which the black hole production is faster than Poincar\'e recurrence. Once such a black hole is formed, it will continue to grow (along with its entropy) and effectively quench the process of recurrence.

In quantum case, the Poincar\'e recurrence is given by the double exponential of the (quantum) entropy, while the black holes can evaporate. We impose an additional condition that the temperature of a newly formed black hole must be lower than the temperature for the gas in order to survive and grow. In Eq.~(\ref{tprq})
we derive the condition which gives us a critical temperature above which the black hole production is faster than quantum Poincar\'e recurrence time. However, in quantum case, the Poincar\'e recurrence theorem can be applied to the black hole states too. Depending on the gas temperature, a black hole can have larger or smaller entropy than the same amount of gas. Therefore, the presence of the black hole can make the Poincar\'e recurrence time longer or shorter. In Eq.~(\ref{tc}), we derive the critical temperature {\it below} which the produced black hole increases the entropy of the whole system (gas particles plus a black hole).

Finally, we emphasized that the condition in Eq.~(\ref{tc}) must be modified if the evolution of the system is fast enough. Namely, black hole initially formed will evolve further by mergers and accretion of surrounding particles. If evolution is fast enough (i.e. faster than the Poincar\'e recurrence time of the gas), then one large surviving black hole will dominate the system. Then we have to compare the total initial entropy of the gas
with the entropy of the final large black hole after accretion.  This condition (given in Eq.~(\ref{tc1})), gives the critical temperature {\it above} which the presence of black holes increase the entropy of the whole system and prolongs the Poincar\'e recurrence time.

In conclusion, the discussion we presented here indicates that the Poincar\'e recurrence phenomenon might be significantly affected by gravity.

\begin{acknowledgments}
The would like to thank Don Page for very useful discussion related to this topic.
This work was partially supported by the US National Science Foundation, under Grant No. PHY-1417317.
\end{acknowledgments}

\appendix

\section{Diluted gas condition}
In the footnote 1, we gave a rough estimate on the initial temperature required to satisfy the condition for diluted gas. Now we carry the calculation to
justify our statements there. The mean diameter of black holes formed during time interval $dt$ can be calculated by inserting $2R_S$ into
 Eq.~(\ref{BHrate0}).
\small
\bea
&&\langle{2R_S}\rangle  =\nonumber\\
&&\frac{\frac12\iint d^3k_1 d^3k_2 2R_S(\vec{k}_1,\vec{k}_2) v_{12}(\vec{k}_1,\vec{k}_2)\sigma_{12}(\vec{k}_1,\vec{k}_2)f(E_1)f(E_2)}{\frac12\iint d^3k_1 d^3k_2 v_{12}(\vec{k}_1,\vec{k}_2)\sigma_{12}(\vec{k}_1,\vec{k}_2)f(E_1)f(E_2)}.\nonumber\\
&=&\frac{4134375\left[\zeta(9/2)\right]^2}{64\pi^7}\frac{Gk_BT}{c^4}.
\eea
\normalsize
We put the details of this integration in App.~(\ref{2R_S}). Note that this result holds for both the classical and quantum case.

For the classical case, we then need $\bar{n}_c\langle{2R_S}\rangle^3<1$, which is satisfied if
\be
\tilde{T}<0.1/\tilde{\bar{n}}_c^{1/3}.
\ee

For the quantum case, the number density of particles is
\bea\label{n_q}
\langle{\bar{n}_q}\rangle&=&\int\frac{d^3k_1}{h^3}f_q(E_1)\nonumber\\
&=&\frac{4\pi g}{h^3}\left(\frac{k_BT}{c}\right)^3\int dx_1\frac{x_1^2}{\exp(x_1)-1}\nonumber\\
&=&\frac{8\pi\zeta(3)g}{h^3}\left(\frac{k_BT}{c}\right)^3,
\eea
where $\zeta(3)$ is the Riemann zeta function. In the last step, we have used Eq.~(\ref{zeta}).

Requiring $\langle{\bar{n}_q}\rangle\langle{2R_S}\rangle^3<1$, we get the constraint which has to be satisfied in order for inter-particle distance in gas to be larger than the Schwarzschild diameter of produced black holes
\be
\tilde{T}<0.3g^{-1/6}.
\ee
This condition is satisfied for any temperature almost up to the Planck scale, thus justifying our assumption.

\section{Calculation of $g(\lambda)$}\label{g(lambda)}
First we evaluate the following integration for later use. For arbitrary $\alpha\ne 1$ and $\beta<1$, then
\bea\label{beta}
&&\iint \text{d$\Omega $}_1 \text{d$\Omega $}_2 (1-\beta  \text{cos$\theta $})^{-\alpha }\nonumber\\
&=&\iint \text{d$\Omega $}_1 \text{d$\Omega $}_2 \left[1+\sum_{n=1}^{\infty } \frac{(-\beta )^{n} \prod _{i=0}^{n-1} (\alpha +i)}{n!}\cos^n\theta\right]\nonumber\\
&=&16\pi^2\left[1+\sum _{m=1}^{\infty } \frac{(-\beta )^{2 m} \prod _{i=0}^{2 m-1} (\alpha +i)}{(2 m+1)!}\right]\nonumber\\
&=&16\pi^2\times\frac{(1-\beta)^{1-\alpha }-(1+\beta )^{1-\alpha}}{2\beta(\alpha -1)}.\nonumber\\
\eea
Here we introduced the parameter $\beta$, because the integral for positive values of $\alpha$ would diverge without it. When evaluating $g(\lambda)$, we will take the limit $\beta\rightarrow 1$
when the potentially diverging terms disappear after some summation. In the second line, we used binomial series expansion.  We can easily see that, expanding
around $\beta=0$ for the last line gives the expansion in the previous line.

In the third line, we used
\bea
B^n&\equiv&\iint d\Omega_1 d\Omega_2\cos^n\theta
=\begin{cases} \frac{16\pi^2}{n+1}, & n~~\text{even},\\
                       0,              & n~~\text{odd},
          \end{cases}
\eea
which is to be derived in App.~(\ref{B^n}).

Therefore,
\small
\bea
&&g(\lambda)\nonumber\\
&\equiv&\iint d\Omega_1d\Omega_2(1-\cos\theta)^{1/4}\exp\big(-\lambda/\sqrt{1-\cos\theta}\big)\nonumber\\
&=&\lim_{\beta\to 1}\iint d\Omega_1d\Omega_2(1-\beta\cos\theta)^{1/4}\exp\big(-\lambda/\sqrt{1-\beta\cos\theta}\big)\nonumber\\
&=&\lim_{\beta\to 1}\iint d\Omega_1d\Omega_2\sum_{n=0}^{\infty}\frac{(-\lambda)^n}{n!}(1-\beta\cos\theta)^{-(\frac{n}2-\frac14)}\nonumber\\
&=&\lim_{\beta\to 1}\sum_{n=0}^{\infty}\frac{8\pi^2(-\lambda)^n}{n!}\frac{(1-\beta)^{\frac54-\frac{n}{2}}-(1+\beta )^{\frac54-\frac{n}{2}}}{\beta(\frac{n}{2}-\frac54)}\nonumber\\
&=&\lim_{\beta\to 1}\frac{32\pi^2}{15\beta}\bigg\{(1+\beta)^{1/4}\Big[4\lambda^2-2\sqrt{1+\beta}\lambda+3(1+\beta)\Big]e^{\frac{-\lambda}{\sqrt{1+\beta}}} \bigg.\nonumber\\
&&~~~~~~~~~~~\bigg. -(1-\beta)^{1/4}\Big[4\lambda^2-2\sqrt{1-\beta}\lambda+3(1-\beta)\Big]e^{\frac{-\lambda}{\sqrt{1-\beta}}} \bigg.\nonumber\\
&&~~~~~~~~\bigg. -4\sqrt{\pi}\lambda^{5/2}\Big[\text{erfc}\Big(\frac{\sqrt{\lambda}}{\sqrt[4]{1+\beta}}\Big) - \text{erfc}\Big(\frac{\sqrt{\lambda}}{\sqrt[4]{1-\beta}}\Big) \Big]\bigg\}\nonumber\\
&=&\frac{2^{25/4}\pi^2}{15} \bigg[\left(2 \lambda ^2-\sqrt{2} \lambda +3\right)\exp\left(-\frac{\lambda }{\sqrt{2}}\right)\bigg.\nonumber\\
&&~~~~~~~~~~~~~~~~~~~\bigg.-2^{3/4} \sqrt{\pi } \lambda ^{5/2} \text{erfc}\bigg(\frac{\sqrt{\lambda }}{\sqrt[4]{2}}\bigg)\bigg].
\eea
\normalsize
Eq.~(\ref{beta}) is used in line 4.

\section{Calculation of $B^n$}\label{B^n}
Recall first
\be
B^n\equiv\iint d\Omega_1 d\Omega_2\cos^n\theta.
\ee
Note that
\be
\cos\theta=\cos\theta_1\cos\theta_2+\sin\theta_1\sin\theta_2\cos(\phi_1-\phi_2),
\ee
where $\theta_{1,2}, \phi_{1,2}$ are the angular components of $\vec{x}_1,\vec{x}_2$ in spherical coordinates, respectively. Then
\bea
B^n&=&\int^{\pi}_0\!d\theta_1\!\int^{\pi}_0\!d\theta_2\!\int^{2\pi}_0\!d\phi_1\!\int^{2\pi}_0\!d\phi_2\sin\theta_1\sin\theta_2\nonumber\\
&&\times\left[\cos\theta_1\cos\theta_2+\sin\theta_1\sin\theta_2\cos(\phi_1-\phi_2)\right]^n\nonumber\\
&=&\sum^n_{m=0}\frac{n!}{m!(n-m)!}\int^{\pi}_0\!d\theta_1\!\int^{\pi}_0\!d\theta_2\!\int^{2\pi}_0\!d\phi_1\!\int^{2\pi}_0\!d\phi_2\nonumber\\
&&\times(\cos\theta_1\cos\theta_2)^{n-m}(\sin\theta_1\sin\theta_2)^{m+1}\cos^m(\phi_1-\phi_2)\nonumber\\
&=&\sum^n_{m=0}\frac{n!}{m!(n-m)!}\left[\int^{\pi}_0d\theta_1(\cos\theta_1)^{n-m}(\sin\theta_1)^{m+1}\right]^2\nonumber\\
&&\times\int^{2\pi}_0\!d\phi_1\!\int^{2\pi}_0\!d\phi_2\cos^m(\phi_1-\phi_2)\nonumber\\
&\equiv&\sum^n_{m=0}\frac{n!}{m!(n-m)!}(B^{nm}_\theta)^2 B^m_\phi.
\eea

$B^m_\phi$ can be calculated using its recursion relations, which is easily derived by integration by parts.
\small
\bea
&&B^m_\phi\nonumber\\
&\equiv&\int^{2\pi}_0d\phi_1\int^{2\pi}_0d\phi_2\cos^m(\phi_1-\phi_2)\nonumber\\
&=&\int^{2\pi}_0d\phi_2\int^{1}_{-1} d\sin(\phi_1-\phi_2)\cos^{m-1}(\phi_1-\phi_2)\nonumber\\
&=&\int^{2\pi}_0\!\!d\phi_2\left[\left(\sin(\phi_1-\phi_2)\cos^{m-1}(\phi_1-\phi_2)\right)\vert^{\phi_1=2\pi}_{\phi_1=0} \right.\nonumber\\
&&\left.+(m-1)\int^{2\pi}_0d\phi_1\sin^2(\phi_1-\phi_2)\cos^{m-2}(\phi_1-\phi_2)\right]\nonumber\\
&=&(m-1){\int^{2\pi}_0\!\!\!d\phi_1\int^{2\pi}_0\!\!\!d\phi_2\left[\cos^{m-2}(\phi_1-\phi_2)-\cos^{m}(\phi_1-\phi_2)\right]}\nonumber\\
&=&(m-1)\left(B^{m-2}_\phi-B^m_\phi\right).
\eea
\normalsize
So we get
\be
B^m_\phi=\frac{m-1}{m}B^{m-2}_\phi.
\ee
$B^0_\phi$ and $B^1_\phi$ are easily calculated,
\be
B^0_\phi=4\pi^2, ~~~~B^1_\phi=0.
\ee
Therefore,
\be
B^m_\phi=\begin{cases} 4\pi^2\frac{(m-1)!!}{m!!}, & m~~\text{even},\\
                       0,                         & m~~\text{odd}.
          \end{cases}
\ee
For even $m$, let $z=\cos\theta_1$, then
\bea
B^{nm}_\theta&\equiv&\int^{\pi}_0d\theta_1(\cos\theta_1)^{n-m}(\sin\theta_1)^{m+1}\nonumber\\
&=&\int^1_{-1}dz z^{n-m}(1-z^2)^{m/2}\nonumber\\
&=&\begin{cases} \frac{\Gamma(1+\frac{m}2)\Gamma(\frac{1-m+n}2)}{\Gamma(\frac{3+n}2)}, & n~~\text{even},\\
                       0,                                                             & n~~\text{odd}.
          \end{cases}
\eea
Here $\Gamma$ is the gamma function.

Thus for even $n$,
\footnotesize
\bea
B^n&=&\sum^n_{\underset{(\text{even})}{m=0}}\frac{n!}{m!(n-m)!}(B^{nm}_\theta)^2 B^m_\phi\nonumber\\
&=&4\pi^2\sum^n_{\underset{(\text{even})}{m=0}}\frac{n!}{m!(n-m)!}\frac{(m-1)!!}{m!!}\left[\frac{\Gamma(1+\frac{m}2)\Gamma(\frac{1-m+n}2)}{\Gamma(\frac{3+n}2)}\right]^2\nonumber\\
&=&4\pi^2\sum^n_{\underset{(\text{even})}{m=0}}\frac{n!(m-1)!!}{m!(n-m)!m!!}\frac{2^{2+m}[(m/2)!]^2[(n-m-1)!!]^2}{[(1+n)!!]^2}\nonumber\\
&=&16\pi^2\sum^n_{\underset{(\text{even})}{m=0}}\frac{n!!(n-1)!!(m-1)!!(m!!)^2[(n-m-1)!!]^2}{(m!!)^2(m-1)!!(n-m)!!(n-m-1)!![(1+n)!!]^2}\nonumber\\
&=&16\pi^2\frac{n!!}{(n+1)(n+1)!!}\sum^n_{\underset{(\text{even})}{m=0}}\frac{(n-m-1)!!}{(n-m)!!}.
\eea
\normalsize
In the following, we shall prove that
\be\label{Bn}
B^n=\begin{cases} \frac{16\pi^2}{n+1}, & n~~\text{even},\\
                       0,              & n~~\text{odd}.
          \end{cases}
\ee
First note directly that
\be
B^0=16\pi^2,~~~~B^2=\frac{16\pi^2}{3}.
\ee
If (\ref{Bn}) holds for $n-2$, that is
\be
\frac{16\pi^2}{n-1}=16\pi^2\frac{(n-2)!!}{(n-1)(n-1)!!}\sum^{n-2}_{\underset{(\text{even})}{m=0}}\frac{(n-m-3)!!}{(n-m-2)!!}.
\ee
So,
\be
\sum^{n-2}_{\underset{(\text{even})}{m=0}}\frac{(n-m-3)!!}{(n-m-2)!!}=\frac{(n-1)!!}{(n-2)!!}.
\ee
Then,
\small
\bea
B^n&=&16\pi^2\frac{n!!}{(n+1)(n+1)!!}\sum^n_{{\underset{(\text{even})}{m=0}}}\frac{(n-m-1)!!}{(n-m)!!}\nonumber\\
&=&16\pi^2\frac{n!!}{(n+1)(n+1)!!}\left[\frac{(n-1)!!}{n!!}+\sum^{n}_{\underset{(\text{even})}{m=2}}\frac{(n-m-1)!!}{(n-m)!!}\right]\nonumber\\
&=&16\pi^2\frac{n!!}{(n+1)(n+1)!!}\left[\frac{(n-1)!!}{n!!}+\sum^{n-2}_{\underset{(\text{even})}{m=0}}\frac{(n-m-3)!!}{(n-m-2)!!}\right]\nonumber\\
&=&16\pi^2\frac{n!!}{(n+1)(n+1)!!}\left[\frac{(n-1)!!}{n!!}+\frac{(n-1)!!}{(n-2)!!}\right]\nonumber\\
&=&\frac{16\pi^2}{n+1}.
\eea
\normalsize
Thus our assumption is proved.

\section{Calculation of $\langle 2R_S\rangle$}\label{2R_S}
Putting the integrals into dimensionless forms, we get
\be
\langle{2R_S}\rangle=\frac{D}{C}\frac{4\sqrt2 Gk_BT}{c^4},
\ee
where $C$ and $D$ are as follows.
\small
\bea
C&\equiv&\iint d^3x_1d^3x_2 \frac{x_1x_2(1-\cos\theta)^{3/2}}{[\exp(x_1)-1][\exp(x_2)-1]},\\
D&\equiv&\iint d^3x_1d^3x_2 \frac{(x_1x_2)^{3/2}(1-\cos\theta)^{2}}{[\exp(x_1)-1][\exp(x_2)-1]}.
\eea
\normalsize
The radial and angular parts are separable. The former is straightforward to integrate, using
\be\label{zeta}
\Gamma(s)\zeta(s)=\int_0^{\infty}\frac{x^{s-1}}{e^x-1}dx.
\ee
Using subscripts $r$ and $a$ to denote the radial and angular parts of the integration, then
\bea
C_r&=&\left[\int_0^{\infty}dx_1\frac{x_1^3}{e^{x_1}-1}\right]^2=\frac{\pi^8}{225},\\
D_r&=&\left[\int_0^{\infty}dx_1\frac{x_1^{7/2}}{e^{x_1}-1}\right]^2=\frac{11025\pi\left[\zeta(9/2)\right]^2}{256}.
\eea

For the angular parts, Eq.~(\ref{beta}) can be used.
\bea
C_a&=&\iint d\Omega_1 d\Omega_2(1-\cos\theta)^{3/2}=\frac{64\sqrt{2}\pi^2}{5},\\
D_a&=&\iint d\Omega_1 d\Omega_2(1-\cos\theta)^{2}=\frac{64\pi^2}{3}.
\eea

Therefore,
\bea
C&=&C_rC_a=\frac{64\sqrt{2}\pi^{10}}{1125},\\
D&=&D_rD_a=\frac{3675\pi^3\left[\zeta(9/2)\right]^2}{4}.
\eea

So
\be
\langle 2R_S\rangle=\frac{4134375\left[\zeta(9/2)\right]^2}{64\pi^7}\frac{Gk_BT}{c^4}.
\ee

\end{document}